\documentclass[sigconf]{acmart}
\usepackage[utf8]{inputenc}

\usepackage{booktabs} 
\usepackage[defblank]{paralist} 
\usepackage{subcaption}
\usepackage{tikz}
\usetikzlibrary{shapes, chains, arrows, fit, arrows.meta}
\tikzstyle{myarrows}=[>=Latex]
\tikzstyle{data3}=[rectangle split,rectangle split parts=3,draw,text centered, text width=2cm]
\tikzstyle{data2}=[rectangle split,rectangle split parts=2,draw,text centered]

\copyrightyear{2018} 
\acmYear{2018} 
\setcopyright{othergov}
\acmConference[PaPoC'18]{5th Workshop on Principles and Practice of Consistency for Distributed Data}{April 23--26, 2018}{Porto, Portugal}
\acmBooktitle{PaPoC'18: 5th Workshop on Principles and Practice of Consistency for Distributed Data, April 23--26, 2018, Porto, Portugal}
\acmPrice{15.00}
\acmDOI{10.1145/3194261.3194262}
\acmISBN{978-1-4503-5655-8/18/04}

\definecolor{gray_ulisses}{gray}{0.55}
\usepackage{listings}
\lstdefinelanguage{repliss}
{
  morekeywords={datatype,atomic,return,in,of,default,initial,if,else,new,forall,exists,
  visible,happened,before,invariant,return,forall,operation, precond, local_precond, assert, read, update,group,generator,effector,stably,let,henceforth},
  sensitive=true,
  morecomment=[l]{\%},
  morecomment=[s]{/*}{*/},
  morestring=[b]",
}
\newcommand{\code}[1]{\mbox{\tt{#1}}}

\begin{document}
\title{
                Ensuring referential integrity under causal
                consistency%
}

\titlenote{Produces the permission block, and
  copyright information}

\author{Marc Shapiro}
\affiliation{Sorbonne Université-LIP6, Paris, France}
\affiliation{Inria, Paris, France}

\author{Annette Bieniusa}
\affiliation{TU Kaiserslautern, Germany}

\author{Peter Zeller}
\affiliation{TU Kaiserslautern, Germany}

\author{Gustavo Petri}
\affiliation{IRIF, Paris Diderot -- Paris 7, France}

\begin{abstract}
  \noindent
  Referential integrity (RI) is an important correctness property of a
  shared, distributed object storage system.
  It is sometimes thought that enforcing RI requires a strong form of
  consistency.
  In this paper, we argue that causal consistency suffices to maintain
  RI\@.
  We support this argument with pseudocode for a \emph{reference} CRDT
  data type that maintains RI under causal consistency.
  QuickCheck has not found any errors in the model.
\end{abstract}

%
%



\maketitle

\section{References and referential integrity}
\label{sec:refer-refer-integr}

Consider a shared store (memory) of objects, and a \emph{reference} data
type for linking objects in the store.
Let's call a referencing object the \emph{source} of the reference, and
the referenced object its \emph{target}.
Intuitively, the \emph{referential integrity} (RI) invariant states that
if an application can reference some target, then the target ``exists,'' in
the sense that the application can access the target safely.
A referenced object must not be deleted; conversely, when an object
cannot be reached by any reference, deleting it is allowed.

We say that an object is \emph{unreachable} if it \emph{is not} the
target of a reference, and \emph{never will be} in the future (the latter
clause is problematic under weak consistency).
The RI property that we wish to achieve is the following:
\begin{compactitem}
\item \emph{Safety:} An object can be deleted only if it is unreachable.
\item \emph{Liveness:} Unreachability of an object will eventually be detected.
\end{compactitem}

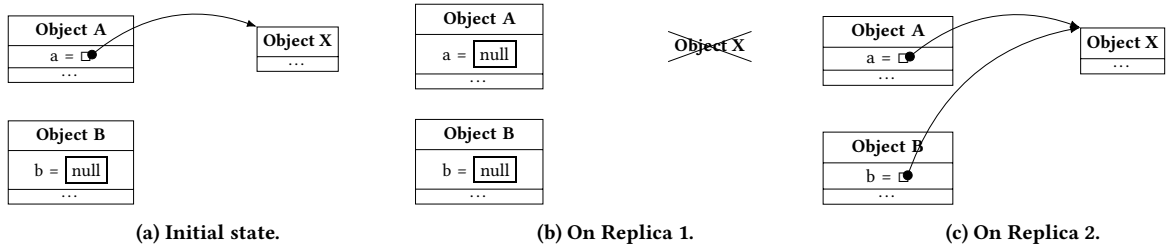
\begin{figure*}
  \centering

\begin{subfigure}[b]{0.3\textwidth}
  \resizebox{!}{0.5\textwidth}{%
\begin{tikzpicture}[node distance=2cm,myarrows]
    \node [data3]                (A) {\textbf{Object A} \nodepart{second} a = $\square$ \nodepart{third} \dots};
    \node [data3, below of=A]    (B) {\textbf{Object B} \nodepart{second} b = \fbox{null} \nodepart{third} \dots};
    \node [data2, right of=A, node distance=4cm]    (C) {\textbf{Object X} \nodepart{second} \dots};
    \path[*->, bend left] (0.31,-0.15) edge (C.north west);
\end{tikzpicture}}
\caption{Initial state.}
\label{fig:init}
\end{subfigure}
\begin{subfigure}[b]{0.3\textwidth}
  \resizebox{!}{0.5\textwidth}{%
\begin{tikzpicture}[node distance=2cm, myarrows]
    \node [data3]                (A) {\textbf{Object A} \nodepart{second} a = \fbox{null} \nodepart{third} \dots};
    \node [data3, below of=A]    (B) {\textbf{Object B} \nodepart{second} b = \fbox{null} \nodepart{third} \dots};
    \node [data2, right of=A, node distance=4cm, cross out]    (C) {\textbf{Object X} \nodepart{second} \dots};
\end{tikzpicture}}
\caption{On Replica 1.}
\label{fig:r1}
\end{subfigure}
\begin{subfigure}[b]{0.3\textwidth}
\resizebox{!}{.5\textwidth}{%
\begin{tikzpicture}[node distance=2cm, myarrows]
    \node [data3]                (A) {\textbf{Object A} \nodepart{second} a = $\square$ \nodepart{third} \dots};
    \node [data3, below of=A]    (B) {\textbf{Object B} \nodepart{second} b = $\square$ \nodepart{third} \dots};
    \node [data2, right of=A, node distance=4cm]    (C) {\textbf{Object X} \nodepart{second} \dots};
    \path[*->,bend left] (0.31,-0.15) edge (C.north west);
    \path[*->,bend left] (0.31,-2.2) edge (C.north west);
\end{tikzpicture}}
\caption{On Replica 2.}
\label{fig:r2}
\end{subfigure}
\caption{Concurrently creating references and deleting objects can lead to dangling references. How should the replicas be reconciled?}\label{fig:concurrent_delete}

\end{figure*}

In a storage system where the application can delete objects explicitly,
the programmer must be careful to preserve the RI invariant.
This problem has been studied in the context of (concurrent) garbage collection for decades.
Folklorically, it is often thought that enforcing RI requires
synchronisation and strong consistency.
In fact, previous work has stated otherwise \cite{gc:mem:rep:sor:1224,
  gc:db:1788, db:syn:1780}.
The main purpose of this paper is to construct a reference data type
demonstrating that causal consistency (with progress guarantees)
suffices to ensure RI and to implement a safe deletion operation.
We support this claim with pseudocode.

The solution that we sketch in this paper uses a form of reference
counting (designed for distributed systems), called \emph{reference
  listing} \cite{sor:nom:1083, gc:mem:rep:sor:1224, sos:gc:mem:1356}.
Objects with a non-empty reference list must not be deleted.

\section{Referential Integrity and causal consistency}
\label{sec:refer-integr-caus}

The safety property of RI is an instance of an implication invariant $P \implies
Q$: \emph{If a reference to an object exists, the object can accessed (has not been deallocated).}
Elementary logic tells us that the sequential pattern of first making
$Q$ true, followed by making $P$ true, will maintain such an invariant
(the ``backward pattern'').
Similarly, making $P$ false followed by making $Q$ false (the ``forward
pattern'') also works.
The backward pattern translates to ``first allocate the object, then assign
reference to it,'' and the forward pattern to ``first delete all
references to object, then delete the object.''

In a concurrent system with causal consistency \cite{syn:mat:1025}, if
two updates are ordered by happened-before \cite{con:rep:615}, then all
processes observe them in the same order.
Therefore, we expect the same patterns to extend to such a system.
Unfortunately, this does not suffice to maintain RI, because
both patterns may be executing in parallel.

It is encouraging to remember that some datatypes can be engineered to
support apparently-conflicting concurrent updates.
For instance, a set can support concurrent insertion and removal of the
same element, by making one operation ``win'' deterministically, the
other one being superseded \cite{syn:rep:sh143}.
However, we cannot re-use this design directly since handling references also requires to handle the referred objects accordingly (including transitive reachability).
Furthermore, while it is easy to ensure safety by never deleting
anything, we also require liveness.

Note that causal consistency is only a safety property; it allows
arbitrarily old versions to be observed.
We need to add a progress guarantee assumption to ensure that our
algorithm is live.


We assume that the objects of interest are accessed only via the
reference datatype discussed herein.
We do not address the more complex problem of objects that are accessible
via some external means, e.g., through a well-known key, through a URL,
or via a database query.
These are called ``root'' objects (in garbage-collection parlance),
which for our purposes are never deleted.

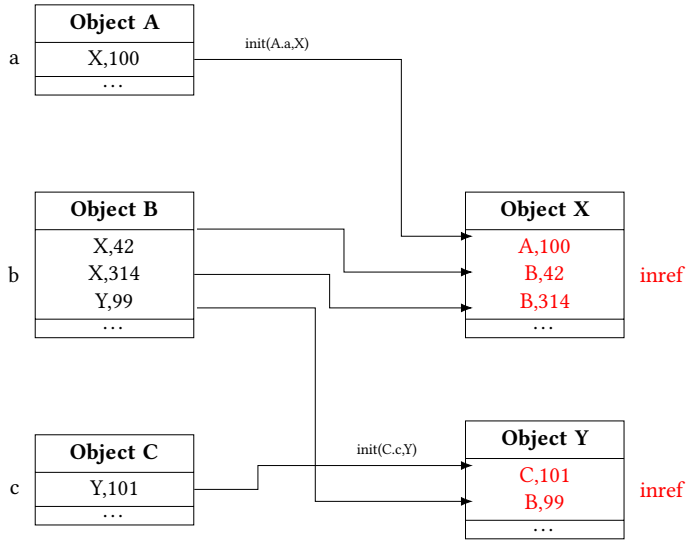
\begin{figure}[tp]
  \centering
  \resizebox{!}{.4\textwidth}{%
  \begin{tikzpicture}[node distance=3cm, myarrows]
      \node [data3]                (A) {\textbf{Object A} \nodepart{second} X,100 \nodepart{third} \dots};
      \node[left=1mm of A.two west] {a};

      \node [data3, below of=A]    (B) {\textbf{Object B} \nodepart{second} X,42 \\ X,314 \\Y,99  \nodepart{third} \dots};
      \node[left=1mm of B.two west] {b};

      \node [data3, below of=B]    (C) {\textbf{Object C}
      \nodepart{second} Y,101  \nodepart{third} \dots};
      \node[left=1mm of C.two west] {c};

      \node [data3, right of=B, node distance=6cm]    (X) {\textbf{Object X}
      \nodepart[color=red]{second} A,100 \\ B,42 \\ B,314  \nodepart{third} \dots};
      \node[color=red,right=1mm of X.two east] {inref};

      \node [data3, right of=C, node distance=6cm]    (Y) {\textbf{Object Y}
      \nodepart[color=red]{second} C,101 \\ B,99 \nodepart{third} \dots};
      \node[color=red,right=1mm of Y.two east] {inref};

      \draw[->] (A.second east) -|  (4,-1.7) node [pos=0.2, above] {{\scriptsize init(A.a,X)}} |- (5,-2.6) ;
      \draw[->] (1.15,-2.5) -| (3.2,-3.0) |- (5,-3.1);
      \draw[->] (B.second east) -| (3,-3.5) |- (5,-3.6);
      \draw[->] (1.15,-3.6) -| (2.8,-4.0) |- (5,-6.3);
      \draw[->] (C.second east) -| (2,-6)  |- (5,-5.8) node [pos=0.8, above] {{\scriptsize init(C.c,Y)}} ;

  \end{tikzpicture}}
  \caption{References with inref/outref after concurrently assigning \texttt{B.b} to \texttt{A.a} (twice) and \texttt{C.c}.}
  \label{fig:ref-example}
\end{figure}

\section{High-level description}
\label{sec:high-level-descr}

We sketch in the following the reference handling protocol; a pseudocode description is given in Appendix A.
A source object contains an instance of a data type called \code{outref} for every attribute that refers to another object.
%
%
A (target) object is associated with exactly one \code{inref}.
The inref identifies the currently-known sources pointing to this target.
Creating a new reference 
 initialises both the inref and an outref.
The only application-level operations supported by inref are
initialisation and testing whether deleting the target is allowed.

A (source) object contains any number of distinct outrefs.
An outref supports the following application-level operations:
\begin{inparaenum}[\it (i)]
\item
  initialisation,
\item
  assigning from another outref,
\item
  assigning \code{null} (we assume that deleting an object first automatically
  nulls out all of its outrefs),
and
\item
  invocation, detailed shortly.
\end{inparaenum}
To support concurrency, assigning an outref behaves much like a
\emph{Multi-Value Register}.
Assignment overwrites its previous value; when concurrent assignments
occur, the resulting reconciled value contains all the
concurrently-assigned values.
To simplify the semantics, we check that the right-hand side of the
assignment is single-valued.

An outref can invoke its target, but this makes sense only if it has a
single (non-null) target.
If the outref contains multiple values, the invocation fails (the
application can fix this by performing a new assignment).

Figure~\ref{fig:ref-example} illustrates three source objects \code{A, B, C},
each containing an attribute single outref named \code{a, b, c}
respectively, and two target objects \code{X} and \code{Y}.
The state illustrated might result from the following code snippet:
\begin{verbatim}
    init (A.a, X); init (C.c, Y);
    B.b := A.a || B.b := A.a || B.b := C.c;
\end{verbatim}


Our algorithm design hinges on two principles that can be implemented
assuming only causal consistency:
\begin{inparaenum}[(1)]
\item
  \emph{before} an outref is assigned to a source object (in initialisation
  or assignment), we ensure that the corresponding inref has been added
  to the target object;
  importantly, causal consistency is enough to enforce
  this ordering of updates.
\item
  To delete a target, we require that no inref exists, nor
  will later be added, for this target.
  This property can be checked by well-known mechanisms which rely only
  on causal consistency and progress guarantees \cite{app:rep:optim:1501}.
\end{inparaenum}
The combination of these properties is sufficient to ensure RI as
defined in the introduction.

\section{System model and pseudocode}
\label{sec:syst-model-pseud}

The pseudocode for references is listed in
Appendix~\ref{sec:pseudocode}.
Some preliminary explanations are required.

References are layered above a lower-level unmanaged addressing
mechanism (similar to a memory address used by the JVM), which we call
\emph{key}; a key uniquely identifies a single discrete (but possibly
replicated) object.

Our system model is based on invocation split into two phases: the
\emph{generator} executes at a single replica, and generates a list of downstream messages
that are eventually received at all replicas and executed by corresponding \emph{effectors}
\cite{syn:rep:sh143,rep:syn:1690, syn:app:sh179}.
At the source replica, the downstream messages are processed atomically with the generator.
Other replicas may observe delays between the different downstream messages,
but they will always receive them in the order specified by the generator.
The generator may check preconditions (noted
\code{precond})  against shared state; if any precondition is
false, the operation fails.
A generator may not have side effects on shared state.
The effector must have the same effect at every replica, and therefore
may not depend on testing shared state.
We assume an operation's preconditions are \emph{stable}, i.e.,
evaluating the precondition to true does not change under any concurrent
operation \cite{syn:app:sh179}.%
\footnote{
  \label{fn:stable-PL}
  The pseudocode also makes use of \code{local\_precond}, which does not
  need to be stable.
  Our use of the term ``stable'' in this section follows the terminology
  used in rely-guarantee logic.
}

We assume causal consistency, i.e., one operation's effector is
delivered (to some replica) only after the effectors of operations that
are visible to it.
We consider two alternatives for composed operations:
\begin{compactitem}
\item
  Atomic: an operation is the atomic composition of all of its sub-operations.
  All the sub-generators (resp.~sub-effectors) compose into a single
  atomic generator (resp.~effector).
  This is somewhat similar to closed-nested transactions, without the
  isolation property.
\item
  Pure causal: An effector updates a single object, but effectors can be
  chained, respecting the order defined in the code.
  This is somewhat similar to transaction chaining.
\end{compactitem}
In both cases,  if any precondition is false, the whole operation fails.
Appendix~\ref{sec:pseudocode} provides pseudocode for the latter
option.%
\footnote{
  The ``atomic'' version is easier to read, but we prefer to
  minimise the assumptions.
  It is obtained from the pure-causal version by replacing the chained
  effectors by a single atomic one with the same text.
}

The logic is relatively simple.
On creating or copying a reference, avoid races by following the
backward direction, first adding to the target, then to the source.
On resetting (removing) a reference, follow the forward direction, first
removing from the source, then from the target.
We deal with concurrency by ensuring every reference has a unique
identifier, and being careful of not losing any information.
The details are tedious, but hopefully explained in the comments.

The \code{may\_delete} operation merits a more detailed explanation.
This operation returns true if and only if the \code{inref} argument is
not reachable; however, in order to break circular reference patterns,
the \code{last\_refs} argument lists references to ignore.
The \code{stably} notation in \code{may\_delete} and in the third
invariant means that the assertion is true, and that there are no
concurrent mutations that could make it false.%
\footnote{
  This is called a ``stable'' property in the literature on distributed
  algorithms; we use ``stably'' to distinguish from the usage in
  Footnote~\ref{fn:stable-PL}.
}
Detecting \code{stably} boils down to detecting termination.
Its implementation is well understood, requiring replicas to know about each
other in order to exchange information on their progress
\cite{app:rep:optim:1501}.
Note that causal consistency is usually defined as a safety guarantee only
\cite{syn:mat:1025,rep:1754}.
In order to ensure that a \code{stably} check eventually succeeds, we
must add an assumption of progress, i.e., reads do not indefinitely
return an old version.

\paragraph{Correctness}
In order to validate the correctness of our CRDT references
implementation, we formalized the system model and pseudocode
implementation in Isabelle/HOL~\cite{Isabelle}
and tested it with Haskell QuickCheck~\cite{QuickCheck}.
The corresponding code is available on GitHub.%
\footnote{
  \url{https://github.com/peterzeller/ref-crdt}
}
The QuickCheck tests generate random executions and then check the first
and the third invariant described in the pseudocode.
To generate interesting random executions, we let each generated event
depend on two randomly chosen previous events.
Then, we randomly decide how many of their effector messages have been
delivered to the new event.
By doing this, it is likely that an event observes other events only
partially, which is a common source of bugs.
Indeed, we were able to discover some flaws in earlier drafts of the
implementation and were able to fix them.
For the updated implementation, our tests did not find a problem after
50\,000 random executions.





\begin{acks}
  {This research is supported in part
    by \grantsponsor{eu}{European H2020}{https://ec.europa.eu/programmes/horizon2020/} project number \grantnum{eu}{732\,505}
    \href{http://LightKone.eu/}{LightKone},
    and
    by the \href{http://RainbowFS.lip6.fr/}{RainbowFS} project
    of \grantsponsor{anr}{Agence Nationale de la Recherche}{http://www.agence-nationale-recherche.fr}, France, number 
    \grantnum{anr}{ANR-16-CE25-0013-01}.
  }
  
\end{acks}

\bibliographystyle{ACM-Reference-Format}
\bibliography{predef,shapiro-bib-ext,shapiro-bib-perso,local} 

\appendix{}

\section{Pseudocode}
\label{sec:pseudocode}

The following pseudocode describes the pure-causal version of references.
The atomic version differs essentially by replacing the cascaded
effectors with a single atomic one.

Block structure is indicated by indentation.
Comments are preceded by the ``\%'' character.

\begin{lstlisting}[language=repliss,numbers=left]
datatype outref of T
   % A reference to an object of type T containing an inref.
   
   % Key of embedding object. If object has multiple outrefs, assume
   % each one has a distinct object_key.
   object_key: write_once register of Key
   % routing information to referenced object
   dest_keys: MV_register of (outkey: Key, id: uid),
                             initial (nullkey, nulluid)

datatype inref
   % key of embedding object
   object_key: write_once register of Key
   % set of reverse references
   rev_refs: 2P_set of (inkey: Key, id: uid),
                              initial emptyset
   % call "init" only once 
   inuse: CRDT_flag, initial false

% if outref exists, inref exists
invariant
   forall r: outref of T
      (k,u) in r . dest_keys ==> (r . object_key, u) in k . inref . rev_refs

% correct type
invariant
   forall r: outref of T
      (k,u) in r . dest_keys ==>  k in T

% once an inref is unreachable, it remains unreachable
invariant
   forall i: inref
      % "stably" = true at all replicas and no concurrent updates in flight
      stably { i . rev_refs = emptyset }
         ==> henceforth { i . rev_refs = emptyset }

%% constructor; not part of API
_create_inref (k: Key, inref: inref)
   % the inref is embedded inside the object with key k
   inref . object_key := k

%% constructor; not part of API
_create_outref (k: Key of T, outref: outref of T)
   % the outref is embedded inside the object with key k
   outref . object_key := k

%% updates outref with new key value; not part of API
_outref_update(outTo: outref of T, new_key: Key)
    generator(outTo, new_key)
        let source_key = outTo . object_key
        let to_reset = outTo . dest_keys . getall ()
        let newuid =  new_uid()
        % explicit effector chaining
        if new_key != nullkey
           effector#1 (outTo, source_key, new_key, to_reset, newuid)
        else
           effector#2 (outTo, source_key, new_key, to_reset, newuid)

    effector#1 (outTo, source_key, new_key, to_reset, newuid)
        % first insert into new target
        new_key . inref . rev_refs . add ((source_key, newuid))
        effector#2 (outTo, source_key, new_key, to_reset, newuid)

    effector#2 (outTo, source_key, new_key, to_reset, newuid)
        % then assign source
        outTo . dest_keys := (new_key, newuid) % conc. assign possible
        forall (k, u) in to_reset
            % chain reset
            effector#3 (k, u, source_key)
    effector#3 (k, u, source_key)
        % finally, remove old reverse refs
        k . inref . rev_refs . remove ((source_key, u))

% create a reference from outref to inref
init (outref: outref of T, inref: inref)
    generator (outref, inref)
        precond ! inref . inuse          % call init only once
        effector#1(inref)
        % run _outref_update effectors after effector#1
        _outref_update(outref, inref.object_key)
    effector#1(inref)
        inref.inuse := true

% Remove an inref.
% Deleting the object that embeds inref calls this; therefore, the
% outer delete will fail if there are any remaining references.
reset (inref: inref)
    generator
        % Non-reachability is monotonic
        precond inref . may_delete()
    effector
        skip

% Remove an outgoing reference
% Deleting the object that embeds the outref calls this.
reset (outref: outref of T)
    generator (outref)
        % same as assigning nullkey:
        _outref_update(outref, nullkey)
 
:= (outTo: outref of T, outVal: outref of T)
   assign (outTo, outVal)

% outTo := outVal
%
% Copy outVal into outTo; reset outTo; in that order.  Either may be
% initially null.  No-op if outVal target already in outTo.
%
% Concurrent "assign"s to outref store multiple values inside MV_register.
% The user should resolve by a subsequent "assign"
%
assign (outTo: outref of T, outVal: outref of T)
    generator (outTo, outVal)
        % simplification: ensure outVal has no more than one target
        % local check, not necessarily stable
        local_precond outVal . dest_keys . count() = 1
        let (new_key, _) = outVal . dest_keys . get1 ()
        _outref_assign(outTo, newKey)

% Use a reference to call the target object
deref (outref: outref of T, invocation: invocation of T)
    generator (outref, invocation)
        % local checks, not necessarily stable
        local_precond outref . dest_keys . count() = 1
        local_precond outref . dest_keys . get1() != (nullkey, _)
        let (key1, _) = outref . dest_keys.get1()
    effector(key1, invocation)
        invoke (key1, invocation)

% Is target object reliably not referenced?
% To be tested in a generator. last_refs: if only these exist we are
% still OK, because the effector will to reset them shortly.
may_delete (inref: inref,
            last_refs: set of outref, default emptyset
           ): boolean
    generator (inref, last_refs)
        % check that the only remaining rev_refs are those in last_refs
        % (none by default)
        last_keypairs: set of (Antidote_key, uid)
                       = { fold (last_refs,
                                lambda (r) cons (r.object_outref, "_")) }
        return stably inref . rev_refs = last_keypairs
    effector ()
        skip

\end{lstlisting}

\end{document}